# Influence of nonuniform magnetization reorientation on spin-orbit torque measurements

Ryan W. Greening[1], Xin Fan[1]

1. Department of Physics and Astronomy, University of Denver, Denver, Colorado 80210, USA

**Abstract**

Measurements of spin-orbit torques in a ferromagnetic/nonmagnetic multilayer are typically based on an assumption that the entire ferromagnetic layer uniformly responds to the spin-orbit torque. This assumption breaks down when the thickness of the ferromagnetic layer is comparable to the dynamic exchange coupling length, which can be as short as a few nanometers in certain measurement geometries. The nonuniform magnetization reorientation coupled with nonuniform contribution of each magnetic sublayer to the magnetoresistance or the Kerr effect may impact the accuracy in the extrapolation of spin-orbit torque, particularly if a thick ferromagnetic layer is used. In this paper, we use numerical models to investigate such an impact in three different techniques: the magneto-optic-Kerr-effect method, the second-harmonic method and the spin torque ferromagnetic resonance method. We show that the second-harmonic and magneto-optic-Kerr-effect methods are prone to be influenced by the nonuniform magnetization reorientation, while the spin torque ferromagnetic resonance method is much less impacted.

## 1. Introduction

In recent years, research on spin-orbit interaction has unveiled numerous phenomena with spin charge interconversions [1-5]. Among these interesting phenomena, spin-orbit torque – a spin torque generated in a magnetic multilayer film by an in-plane electric current via the spin-orbit interaction – has attracted considerable attention. Due to the high efficiency in switching magnetization [6, 7], moving magnetic domains and skyrmions [8-10], the spin-orbit torque holds potential in the development of future magnetic random access memories [11]. Various measurement techniques have been developed to quantify the spin-orbit torque, many of which are based on the perturbation method. Typically, a small applied in-plane electric current generates a spin-orbit torque that reorients the magnetization of the magnetic layer. The magnetization reorientation is detected through measurements of the magnetoresistance, Hall effect or the Kerr effect [12-19]. In these measurements, it is usually assumed that the magnetization reorientation due to the spin-orbit torque is uniform throughout the magnetic film. However, in a typical spin-orbit device consisting of a FM/NM bilayer, where the FM is a ferromagnetic material and the NM is a nonmagnetic material, the spin-orbit torque is only generated at the two surfaces of the FM due to strong spin dephasing [20]. If the FM thickness is comparable to its dynamic exchange coupling length, the magnetization reorientation due to the spin-orbit torque is not necessarily uniform across the film thickness. On the other hand, the calibration procedures in these experiments are often conducted by applying a magnetic field that rotates the magnetization uniformly. It is therefore important to take into consideration the nonuniform magnetization reorientation when quantifying the magnitude of the spin-orbit torque.

## 2. Model

In this paper we use a FM/NM bilayer with in-plane magnetization as an example. An in-plane current can generate spin-orbit torques, possibly from the spin Hall effect of the NM, the spin-orbit coupling at the FM/NM interface or the transverse spin Hall effect of the FM itself [21]. Regardless of the microscopic origin, due to the strong spin dephasing [20, 22], spin torques shall arise only at the top and bottom surfaces of the FM layer, denoted as $\tau_T$ and $\tau_B$, respectively [23], as illustrated in Fig. 1(a). While the torques should only generate effective fields on the FM layers near the surfaces, the resultant magnetization tilting propagates through the FM bulk via exchange coupling. The propagation length, which is referred to as the dynamic exchange coupling length here, depends on the exchange constant as well as the susceptibility of the FM.

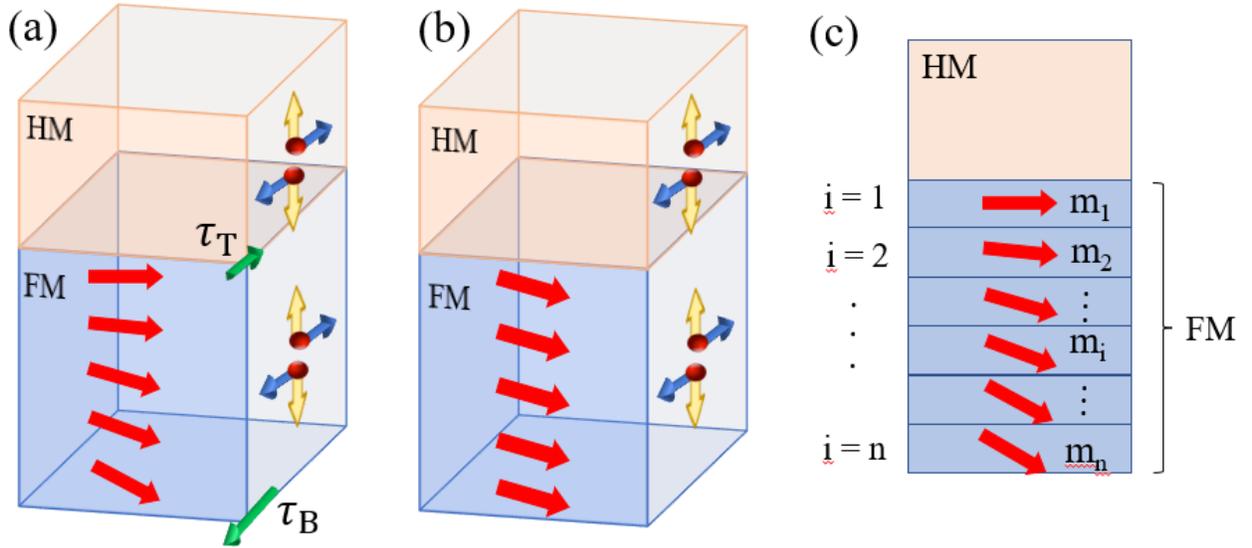

Figure 1 Illustration of magnetization response to surface spin torques for (a) non-uniform magnetization reorientation under finite exchange coupling and (b) uniform magnetization reorientation under infinite exchange coupling. Here red arrows are magnetization vectors. Blue arrows are spin orientations, yellow arrows represent spin current flow directions and green arrows denote spin torques at the top and bottom interfaces. (c) Illustration of how the ferromagnetic layer is divided into sublayers for numerical simulations in this paper.

### 2.1 Dynamic exchange coupling length

We model the ferromagnet by dividing it into many thin layers, each with thickness of *a*. The magnetization is initially uniform across the film. Driven by surface spin torques, the non-uniform magnetization tilting gives rise to an effective exchange coupling field from nearest neighbor layers,

$$\mathbf{h}_{\text{exc}} = \frac{2A_{\text{exc}}}{\mu_0 M_s a^2}[(\mathbf{m}_{i-1} - \mathbf{m}_i) + (\mathbf{m}_{i+1} - \mathbf{m}_i)] = \frac{2A_{\text{exc}}}{\mu_0 M_s}\frac{d^2 \delta \mathbf{m}_i}{dz^2} \qquad (1)$$

where $A_{\text{exc}}$ is the exchange stiffness, $M_s$ is the saturation magnetization, $\mathbf{m}$ is the unit magnetization vector with subscript as the layer index, $\delta \mathbf{m}_i$ is the deviation of magnetization from its initial equilibrium direction.

Therefore, for FM layers within the bulk, the magnetization reorientation is only due to the effective exchange coupling field,

$$\delta \mathbf{m}_i = \frac{\chi \mathbf{h}_{\text{exc}}}{M_s} = \frac{2\chi A_{\text{exc}}}{\mu_0 M_s^2} \frac{d^2 \delta \mathbf{m}_i}{dz^2} \tag{2}$$

, where $\chi$ is the susceptibility.

For the top surface layer, the magnetization reorientation is due to the combination of effective exchange coupling field as well as the effective field due to the spin torque.

$$\delta \mathbf{m}_1 = \frac{\chi}{M_s}\left[\frac{2A_{\text{exc}}}{\mu_0 M_s a^2}(\mathbf{m}_2 - \mathbf{m}_1) + \frac{\tau_T}{\mu_0 M_s a}\right]. \tag{3}$$

When we approximate $a \to 0$, Eq. (3) can be simplified to

$$\frac{d\delta \mathbf{m}_1}{dz} = \frac{\tau_T}{2A_{\text{exc}}} \tag{4}$$

Combining Eqs. (2) and (4), we can derive the magnetization reorientation due to surface torques $\tau_T$ and $\tau_B$ as

$$\delta \mathbf{m}(z) = \chi \frac{\tau_T \cosh\frac{z}{\lambda} + \tau_B \cosh\frac{d-z}{\lambda}}{\mu_0 M_s^2 \lambda \sinh\frac{d}{\lambda}}, \tag{5}$$

Where $d$ is the magnetic film thickness, and $\lambda = \sqrt{\frac{2\chi A_{\text{exc}}}{\mu_0 M_s^2}}$ is the dynamic exchange coupling length. We assumed $\chi$ as a scalar in the above derivation, however in the case of ferromagnetic resonance, $\chi$ should be expressed as a 2 x 2 matrix. The equations (2-5) are still applicable when $\chi$ and $\lambda$ are treated as matrices, $\tau_T$ and $\tau_B$ are treated as vectors. All relevant matrices in Eq. (5) are commutative.

It is worth noting that the dynamic exchange coupling length is different from the magnetostatic exchange length $\sqrt{\frac{2A_{\text{exc}}}{\mu_0 M_s^2}}$ [24], which is mainly for describing domain walls and only depends on the exchange stiffness. Instead, the dynamic exchange coupling length $\lambda$ in our study also depends on the magnetic susceptibility. This can be understood from Eq. (2) that the influence on magnetization from neighboring layers increases with susceptibility, and therefore propagates further. As a result, the dynamic exchange coupling length varies depending on the measurement geometry and even measurement technique. Taking a Py/Pt bilayer with in-plane magnetization as an example, the damping-like torque is equivalent to an out-of-plane magnetic field that tilts the magnetization out of plane. In a DC measurement, the corresponding susceptibility is $\chi = \frac{M_s}{|H_{\text{ext}}| + M_{\text{eff}}}$, where $H_{\text{ext}}$ is the in-plane applied magnetic field, $M_{\text{eff}}$ is the effective demagnetizing field. Using $A_{\text{exc}} = 10^{-11} \text{J} \cdot \text{m}$, $\mu_0 H_{\text{ext}} = 0.01$ T $\mu_0 M_s = 1$ T, we estimate $\chi \approx 1$ and $\lambda \approx 5$ nm. However, for field-like torque that rotates magnetization in the film plane, the relevant susceptibility is $\chi = \frac{M_s}{|H_{\text{ext}}|} = 100$. The dynamic exchange coupling length for this in-plane magnetization rotation is estimated to be about 50 nm. In a spin-torque ferromagnetic resonance measurement, $\chi$ is much enhanced by resonance, thus leading to a long dynamic exchange coupling length. Therefore, as we will see in a numerical simulation later, the magnetization distribution in a spin-torque ferromagnetic resonance measurement is much more uniform than that in a DC/AC measurement.

## 2.2 Simulation of spin-orbit torque measurements responses

In this numerical study, without loss of generality, we will use Py/NM bilayer as an example and compare results from two different Py thicknesses: 3 nm and 8 nm. The 3 nm Py thickness is shorter than the dynamic exchange coupling length in all measurements, while the 8 nm Py may be thicker than the dynamic exchange coupling length in some measurements. For each sample, we will assume two scenarios for the surfaces spin-orbit torques: (1) $|\tau_T| < |\tau_B|$ with an exemplary case of Py/Pt [23] (2) $|\tau_T| > |\tau_B|$ with an exemplary case of Py/Ta.

The damping-like torque measurement results of these samples will be simulated via three different techniques: (1) the magneto-optic-Kerr-effect (MOKE) based spin-orbit torque measurement, (2) the second-harmonic measurement, and (3) the spin-torque ferromagnetic resonance measurement. We will first present the magnetization reorientation by damping-like torque in these measurements. The resultant MOKE and electrical signals will be simulated and compared to those under the assumption of uniform magnetization reorientation.

### 2.2.1 Simulation of the MOKE-based measurement

In the MOKE-based spin-orbit torque measurement [17], an in-plane electric current generates a damping-like torque, that tilts magnetization out of plane. The out-of-plane magnetization is detected via the polar MOKE response. The polar MOKE signal due to the damping-like torque is expected to resemble magnetization hysteresis. Therefore, the applied external magnetic field can be small, e.g. $\mu_0 H_{\text{ext}} = 0.01$T. The dynamic exchange coupling length is calculated to be about 5 nm. Under the influence of damping-like torques $\tau_T$ and $\tau_B$ at the surfaces, we simulate the out-of-plane magnetization tilting $m_z$. As a comparison, we also simulate an extreme case with infinite exchange stiffness $A_{\text{exc}} \to \infty$, where magnetization uniformly tilts $m_z^0$ responding to the total torque, $\tau_T + \tau_B$. In this case, the uniform magnetization reorientation can be calculated as $m_z^0 = \frac{\tau_T + \tau_B}{\mu_0 M_s^2 d}$. In Fig. 2, we plot the deviation of $m_z$ from $m_z^0$, $\Delta m_z = m_z - m_z^0$, normalized by $m_z^0$, as a function of layer position $z$. The four plots correspond to combinational conditions of two different Py thicknesses (3 nm and 8 nm) and two different surface torque ratios ($\tau_T/\tau_B = 0.2$ and $\tau_T/\tau_B = $ -1.3). For 3 nm Py, which is thinner than the dynamic exchange coupling length, the magnetization reorientation is relatively uniform. On the other hand, the nonuniform magnetization reorientation in the 8 nm Py is much more pronounced. The $\tau_T/\tau_B = $ -1.3 case promotes stronger nonuniform magnetization reorientation than the $\tau_T/\tau_B = 0.2$ case, because the former has relatively weaker total spin-orbit torque compared to individual surface torque.

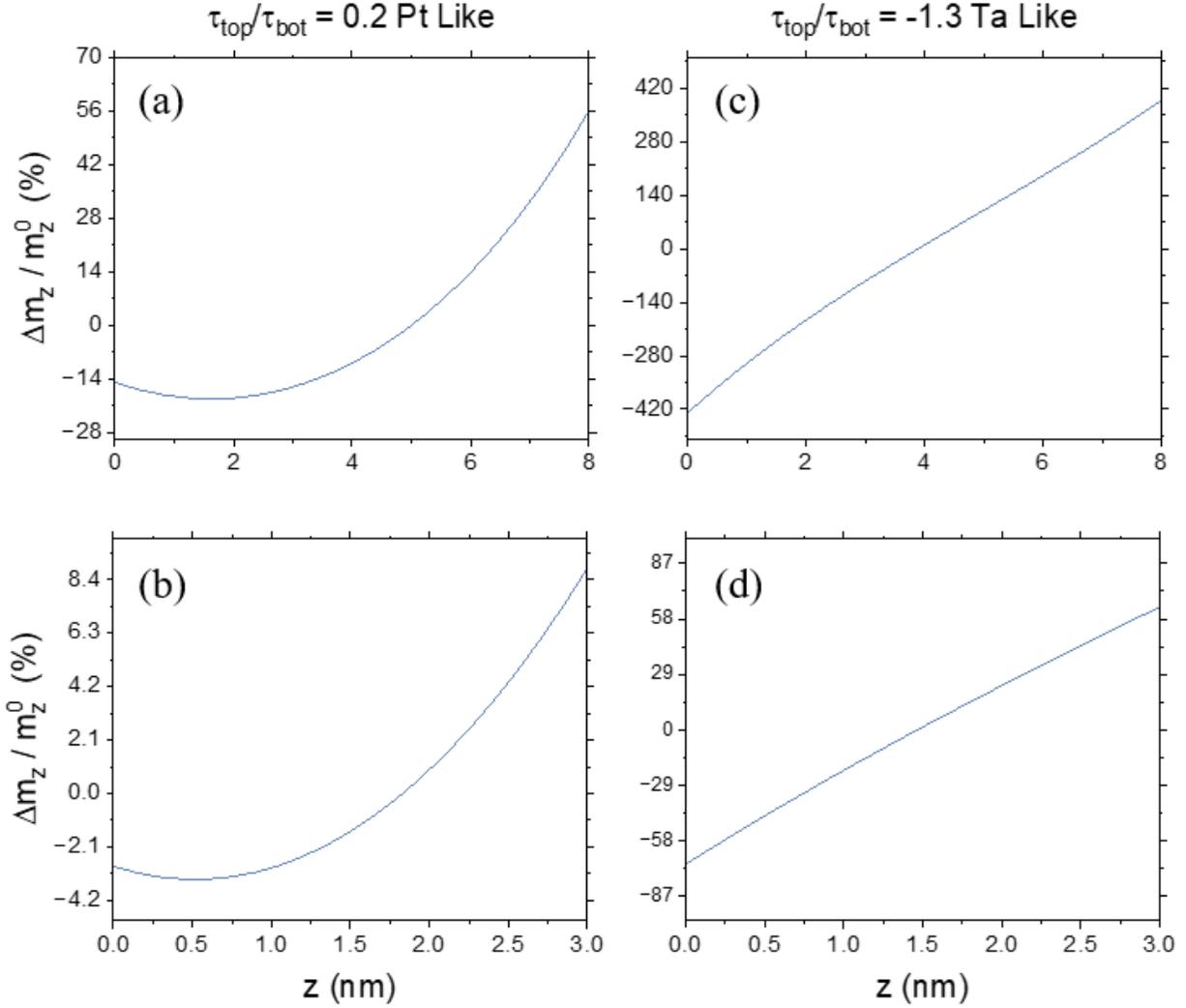

Figure 2 Simulation of magnetization reorientation due to surface spin-orbit torques for 8 nm Py ((a) and (c)) and 3 nm Py ((b) and (d)) in a typical MOKE measurement. Two types of torques are considered: $\tau_T/\tau_B = 0.2$, where the heavy metal has a spin Hall angle like Pt for cases (a) and (b), and $\tau_T/\tau_B = -1.3$, where the heavy metal has a spin Hall angle like Ta for cases (c) and (d).

Since light has a finite penetration depth in typical metals, the MOKE response of each magnetic sublayer is not a constant. To the first order approximation, the total polar MOKE signal can be expressed as $\theta_k = \int_0^d \vartheta(z) m_z(z)\, dz$, where $\vartheta(z)$ is the MOKE contribution from the layer at height $z$, $m_z(z)$ is the out-of-plane magnetization rotation due to the spin torques. We model the MOKE response using the propagation matrix method [25]. Detailed simulation and parameters used are discussed in Appendix A1. Due to the wave nature of light, the Kerr angle is generally a complex number. Figure 3(a) and (b) shows the real and imaginary part of $\vartheta(z)$ as a function of layer height $z$ for 8 nm Py and 3 nm cases respectively. There are considerable variations of $\vartheta(z)$ as a function of $z$. The variation of $\vartheta(z)$ couples with the non-uniform magnetization reorientation $m_z(z)$ will lead to a different overall MOKE signal from what is expected in a uniform magnetization reorientation (assuming $A_{exc} \to \infty$). In Figs.3 (c) and (e), we plot the real part of the MOKE signals as a function of Py thickness $d$ for both the realistic non-uniform magnetization reorientation and the assumed uniform magnetization reorientation in two surface torques scenarios ($\tau_T/\tau_B = 0.2$ and $\tau_T/\tau_B = -1.3$). Figures 3(d) and (f) display the corresponding percentage difference in MOKE response between the uniform and non-uniform magnetization reorientation cases, $\varepsilon_{\text{MOKE}} = \frac{\theta_{\text{Kerr}}^{\text{non-uniform}} - \theta_{\text{Kerr}}^{\text{uniform}}}{\theta_{\text{Kerr}}^{\text{uniform}}} \times 100\%$. The deviation is very small when Py is comparable to the dynamic exchange coupling length (~ 5 nm), but increases dramatically as Py gets thicker than 15 nm due to the combination of non-uniform magnetization reorientation as well as the finite light penetration depth.

Based on these simulation results, we conclude that when using MOKE to measure the net spin-orbit torque, the thickness of the ferromagnetic metal should be chosen to be shorter or comparable to the dynamic exchange coupling length. In this case, the deviation from the assumed uniform magnetization reorientation will be negligibly small. On the other hand, if only one surface spin torque is of interest, one can choose the ferromagnet to be considerably thicker than the dynamic exchange coupling length and light penetration depth to, for example, observe the anomalous spin-orbit torque in a single-layer ferromagnet [23].

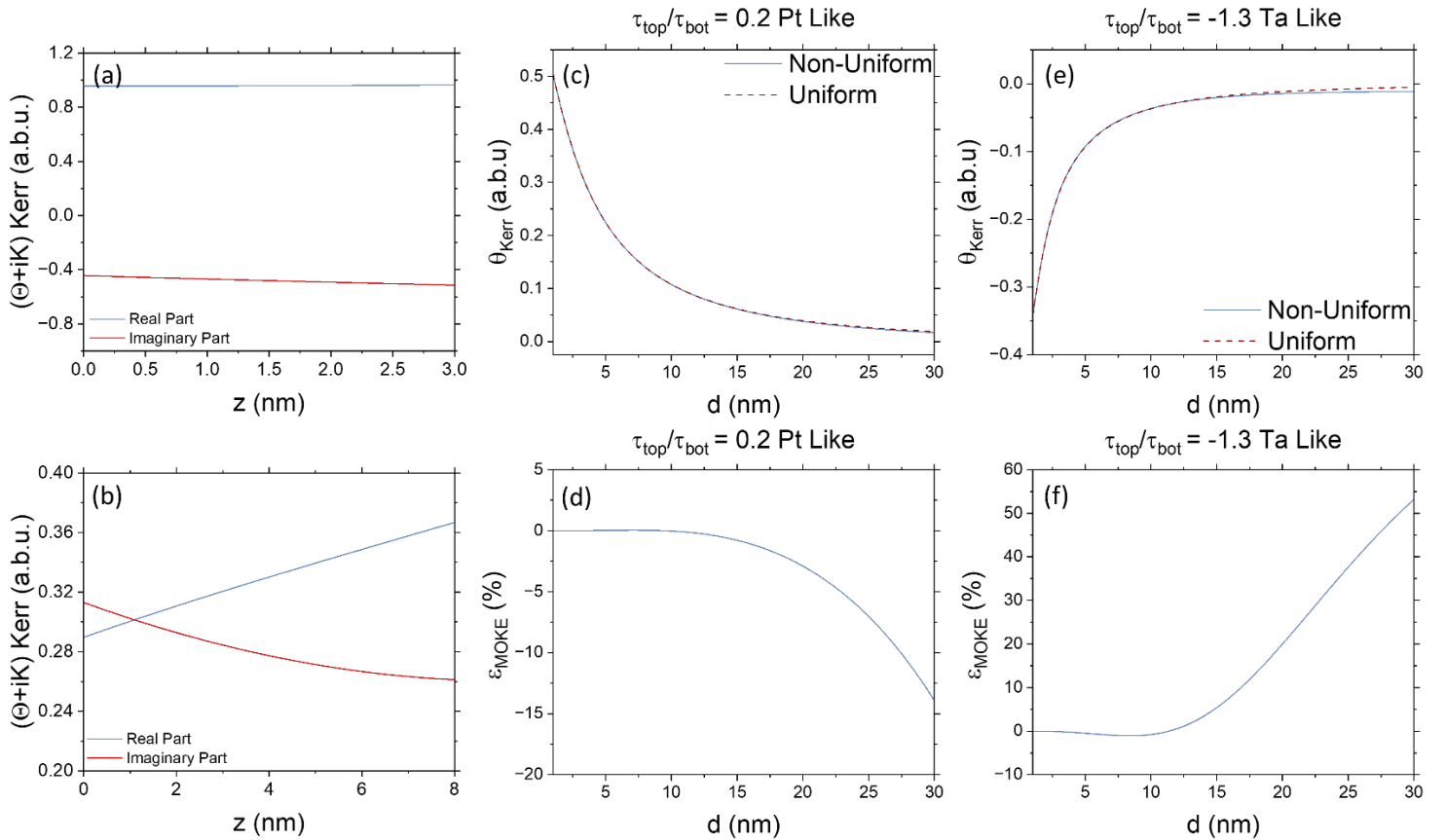

Figure 3 MOKE response simulation. (a-b) Kerr angle contribution from each magnetic layer as a function of $z$-position for 8 nm (a) and 3 nm (b) Py. (c,e) Real part of the MOKE response under uniform and nonuniform magnetization tilting with $\tau_T/\tau_B = 0.2$ (c) and $\tau_T/\tau_B = -1.3$ (e) as a function of film thickness. (d,f) Percent difference in MOKE response ($\varepsilon_{\text{MOKE}}$) between the uniform and non-uniform magnetization reorientation.

### 2.2.2 Simulation of the second-harmonic measurement

For the second-harmonic measurement of the damping-like torque on in-plane magnetized film, we adopt the protocol developed by Avci *et al.* [26]. An in-plane electric current is applied through the sample and a second-harmonic transverse voltage is measured. The second-harmonic voltage signal consists of the planar Hall voltage due to field-like torque, the anomalous Hall voltage due to damping-like torque and the anomalous Nernst effect due to Joule's heating. The three signals can be distinguished via external magnetic field dependence. To suppress the planar Hall voltage contribution, a large in-plane magnetic field shall be applied. In this simulation, we choose $\mu_0 H_{ext} = 1$T. The dynamic exchange coupling length is about 3.5 nm. The magnetization reorientations, as shown in Fig. 4 have very similar behavior as those in Fig. 2, but with larger deviation from uniform magnetization due to even shorter dynamic exchange coupling length.

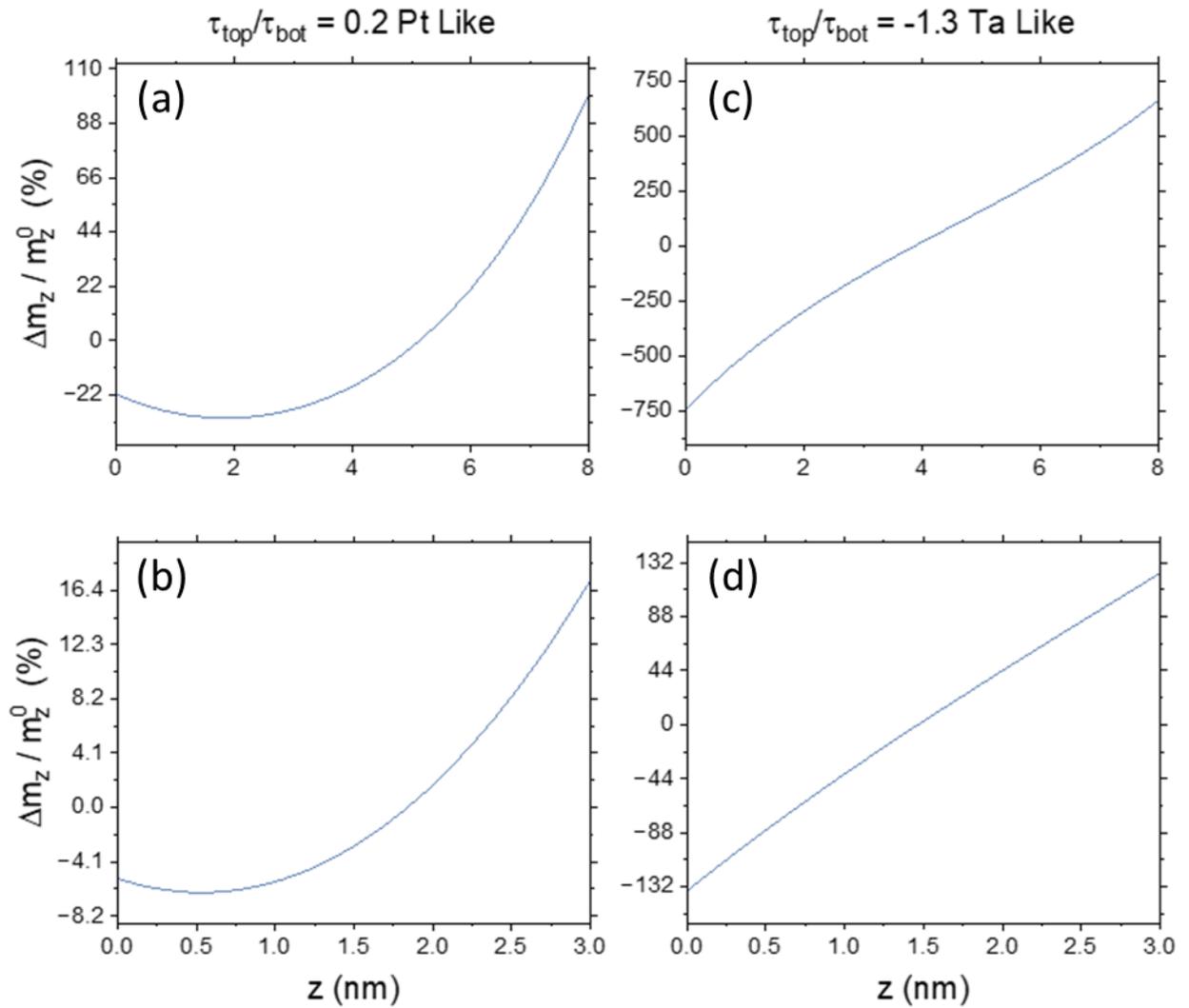

Figure 4 Simulation of magnetization reorientation due to surface spin-orbit torques for 8 nm Py ((a) and (c)) and 3 nm Py ((b) and (d)) in a typical second-harmonic measurement. Two types of torques are considered: $\tau_T/\tau_B = 0.2$,

where the heavy metal has a spin Hall angle like Pt for cases (a) and (b), and $\tau_T/\tau_B$ = -1.3, where the heavy metal has a spin Hall angle like Ta for cases (c) and (d).

The second-harmonic anomalous Hall voltage can be expressed as $V_{AH} = \int_0^d R j_e(z) \theta_{AH}(z) m_z(z) \, dz$, where $R$ is the total resistance in the transverse direction, $j_e(z)$ is the electric current density, and $\theta_{AH}(z)$ is the anomalous Hall angle. The electric current density $j_e(z)$ is $z$-dependent because of interface scattering. Generally speaking the current density is lower near the surfaces, and higher in the center of the film. The scattering rate depends on the details of the interfaces. For simplicity, we will assume that the NM in the Py/NM multilayer has the same electricity as Py, and ignores the scattering at the Py/NM interface. In the simulation of the electric current density distribution, as illustrated in the inset of Fig. 5(a), we treat the Py/NM as a single "extended" Py film with the thickness equals to the total thickness of the original heterostructure, $\tilde{d} = d + d_{NM}$, where $d_{NM}$ is the NM thickness. Only the electric current within $0 \leq z \leq d$ contributes to the anomalous Hall signal. We use the Fuchs-Sondheimer method [27] and assume infinite scattering at both surfaces of the "extended" Py film. The thickness-dependent current density distribution is given as

$$j[z] \propto \int_0^{\pi/2} \sin^3 \beta \left[1 - e^{\frac{-\tilde{d}/2}{\lambda_{mf} \cos \beta}} \cosh\left[\frac{z - \tilde{d}/2}{\lambda_{mf} \cos \beta}\right]\right] d\beta, \tag{6}$$

where $\lambda_{mf}$ is the mean free path of the "extended" Py film. Using $\lambda_{mf} = 5$ nm, $d_{NM} = 3$ nm, we simulate the distribution of the current density as in Fig. 5 (a-b), where asymmetric current distribution is observed. Assuming there is no distribution of the anomalous Hall angle, i.e. $\theta_{AH}(z) = constant$, the second-harmonic anomalous Hall voltage can be simulated as a function of Py thickness $d$. As shown in Figs 5 (c,e), deviations are observed between the case of nonuniform magnetization tilting (the one taking into account the finite dynamic exchange length) and the case of uniform magnetization tilting (assuming infinite exchange coupling). Figures 5(d) and (f) display the corresponding percentage difference in the anomalous Hall response between the uniform and non-uniform magnetization reorientation cases, $\varepsilon_{AH} = \frac{V_{AH}^{non-uniform} - V_{AH}^{uniform}}{V_{AH}^{uniform}} \times 100\%$. The deviation becomes prominent when the Py thickness is a few times the dynamic exchange length, suggesting the uniform magnetization tilting assumption may yield a large error in the extrapolation of the spin-orbit torque when the ferromagnetic layer is relatively thick.

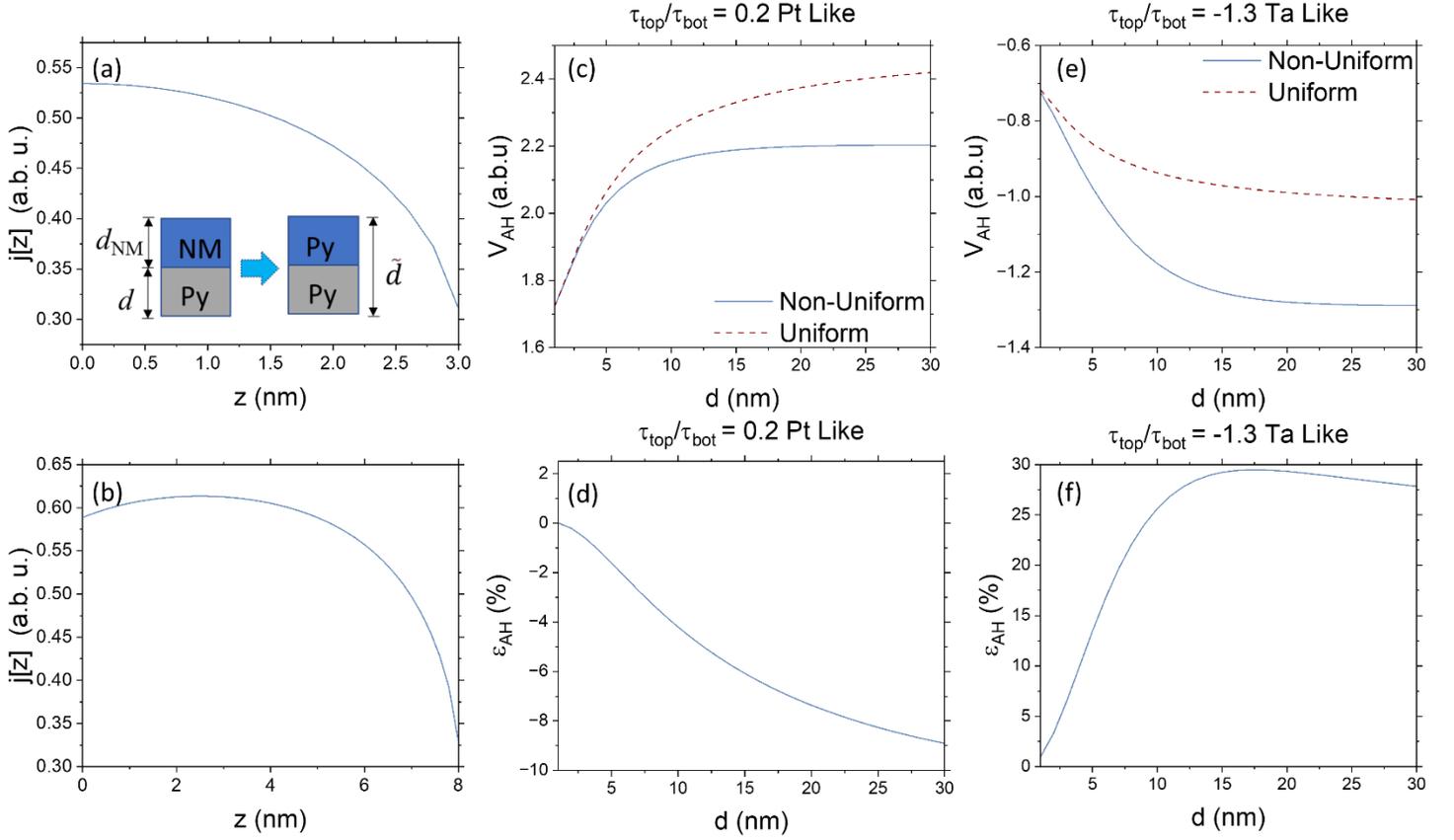

Figure 5 Simulation of the second-order anomalous Hall voltage response. (a-b) Current distribution within Py for 8 nm and 3nm Py samples. Inset of Fig. (a) illustrates the approximation taken for this simulation. The Py/NM bilayer is treated as an extended single Py layer. Here $d_{NM}$ is chosen to be 3 nm. (c,e) Simulated anomalous Hall voltage under uniform and nonuniform magnetization tilting with $\tau_T/\tau_B = 0.2$ (c) and $\tau_T/\tau_B = -1.3$ (e) as a function of film thickness. (d,f) Percent difference in anomalous Hall voltage response ($\varepsilon_{AH}$) between the uniform and non-uniform magnetization reorientation.

It should be emphasized that the simulation of the current density is oversimplified compared to a realistic situation. However, we believe it conveys a qualitatively accurate picture that a nonuniform current distribution coupled with a nonuniform magnetization tilting will lead to an overall anomalous Hall voltage different from what one might expect under uniform magnetization tilting. In addition, the assumption of a constant anomalous Hall angle may also be challenged. An anomalous Hall-like signal can also arise from the spin Hall magnetoresistance and the imaginary part of the spin mixing conductance [28, 29]. This mechanism will give rise to an additional anomalous Hall angle that only depends on the magnetization at the Py/NM interface.

### 2.2.3 Simulation of the spin-torque ferromagnetic resonance measurement

For the spin-torque ferromagnetic resonance measurement [12], a rf current is applied through the sample. The spin-orbit torque drives in-plane magnetization precession, resulting in an oscillation of the anisotropic magnetoresistance. The resistance oscillation couples with the rf current, giving rise to a rectifying voltage. The voltage signal that resembles a symmetric Lorentzian-dependence on external field is attributed to the damping-like torque. In this simulation, we choose the rf frequency to be 6 GHz, with an external magnetic field $\mu_0 H_{ext} \approx 0.05$ T, slightly off resonance. When other magnetic field values near resonance are chosen, we observe qualitatively the same results. We first simulate the in-plane magnetization precession as a result of damping-like torques at the surfaces. Shown in Fig. 6, the non-uniformity of magnetization precession in both 3 nm Py and 8 nm Py sample is negligibly small. This can be explained by the long dynamic exchange length, enhanced by the large susceptibility at resonance.

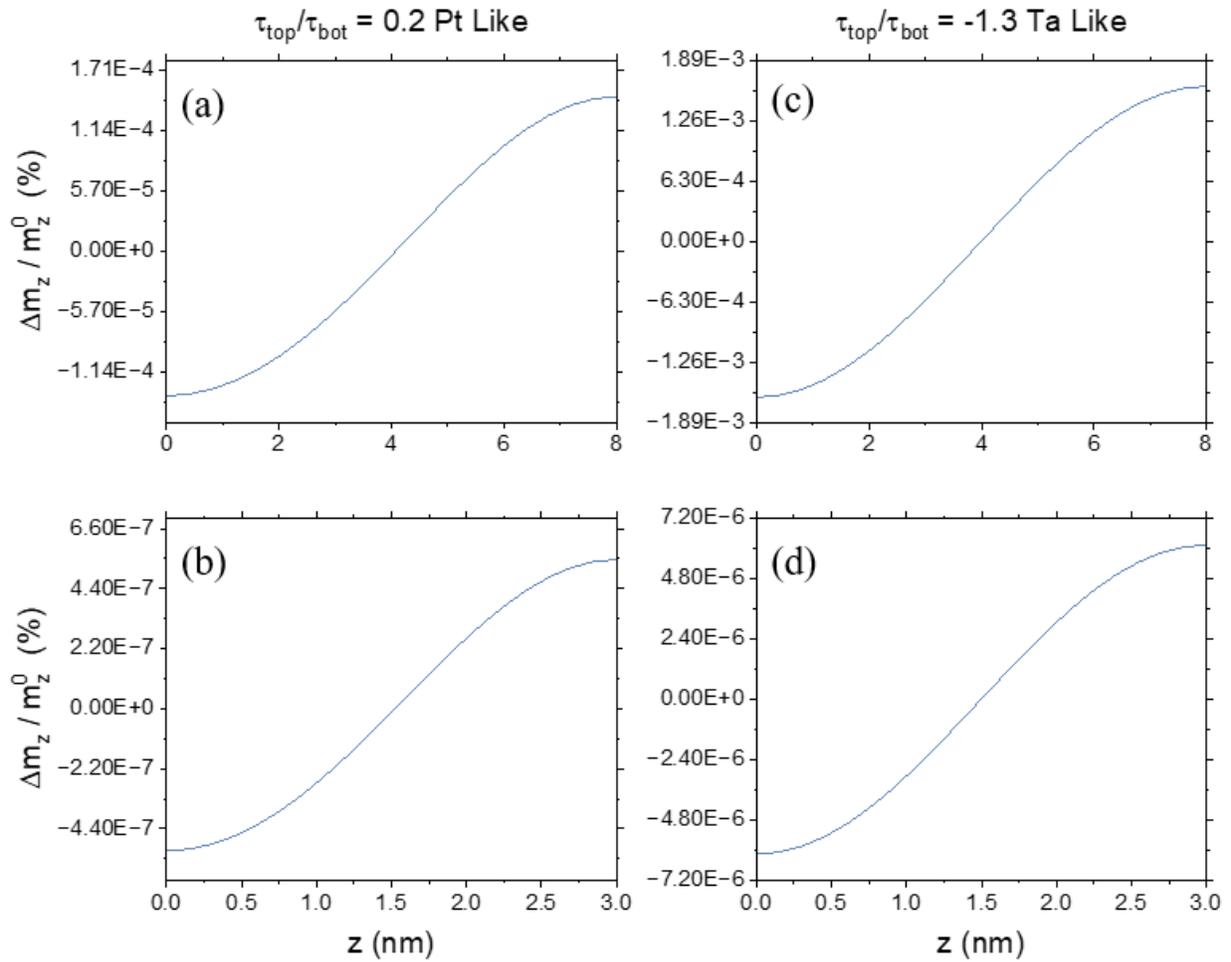

Figure 6 Simulation of magnetization reorientation due to surface spin-orbit torques for 8 nm Py ((a) and (c)) and 3 nm Py ((b) and (d)) in a typical spin torque ferromagnetic resonance measurement. Two types of torques are considered: $\tau_T/\tau_B = 0.2$ for cases (a) and (b), and $\tau_T/\tau_B = -1.3$ for cases (c) and (d).

The spin-torque ferromagnetic resonance measures the rectifying voltage due to the anisotropic magnetoresistance, which can be expressed as $V_{\text{ST-FMR}} = \int_0^d R j_e(z) \theta_{\text{AMR}}(z) \Delta m_x(z)\, dz$, where $R$ is the total resistance in the transverse direction, $j_e(z)$ is the electric current density, and $\theta_{\text{AMR}}(z)$ is the anisotropic magnetoresistance ratio, $\Delta m_x(z)$ is the in-plane magnetization reorientation due to the spin-orbit torques. We assume the same layer-dependent current density $j_e(z)$ as that in section 2.2.2. Figure 7 (a,c) shows the rectified $V_{\text{ST-FMR}}$ signals for both non-uniform and uniform magnetization reorientations in the two different spin torque configurations. Figures 7(b) and (d) display the corresponding percentage difference in the rectifying voltage between the uniform and non-uniform magnetization reorientation cases, $\varepsilon_{\text{ST-FMR}} = \frac{V_{\text{ST-FMR}}^{\text{non-uniform}} - V_{\text{ST-FMR}}^{\text{uniform}}}{V_{\text{ST-FMR}}^{\text{uniform}}} \times 100\%$. Because there is nearly no layer-dependent magnetization reorientation, the deviation of $V_{\text{ST-FMR}}$ due to nonuniform magnetization tilting from that due to uniform magnetization tilting is negligibly small even for relatively thick Py films.

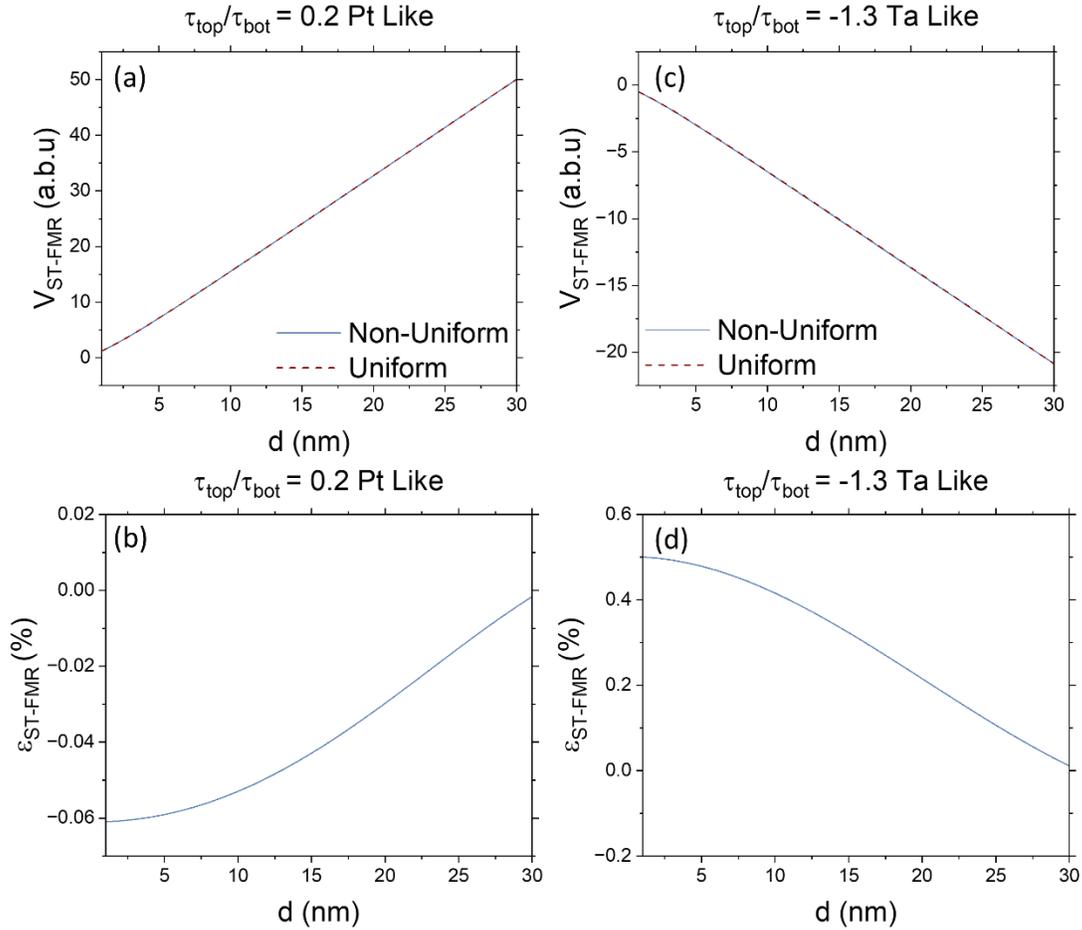

Figure 7 Simulation of the ST-FMR voltage response. (a,c) Rectified $V_{\text{ST-FMR}}$ signals for both non-uniform and uniform magnetization reorientations tilting with $\tau_T/\tau_B = 0.2$ (a) and $\tau_T/\tau_B = -1.3$ (b) as a function of film thickness. (b,d) Percent difference in ST-FMR voltage response ($\varepsilon_{ST-FMR}$) between the uniform and non-uniform magnetization reorientation.

## 3. Discussion

Since the spin-orbit torque arises where inversion symmetry is broken, in a typical ferromagnetic/nonmagnetic bilayer film, the spin-orbit torques are usually located at the two surfaces of the ferromagnet. The influence of the spin-orbit torque propagates within the ferromagnet via exchange coupling, but not indefinitely. The dynamic exchange coupling length depends not only on the exchange constant of the ferromagnet, but also the magnetic susceptibility of the specific measurement technique. When the ferromagnet is thinner than the dynamic exchange coupling length, it is a reasonable approximation that the magnetization uniformly responds to the total spin-orbit torques. However, when the ferromagnet thickness exceeds the dynamic exchange coupling length, magnetization reorientation is nonuniform. The nonuniform magnetization tilting coupled with nonuniform response from different probing method may give rise to a signal that is different from what is commonly anticipated from a uniform magnetization reorientation.

Of the three techniques that we studied, the spin-torque ferromagnetic resonance generally yields the longest dynamic exchange coupling length, making it more suitable for measuring total spin-orbit torques in thick ferromagnets. The second-harmonic method with the smallest magnetic susceptibility tends to have the shortest dynamic exchange coupling length. Moreover, since the magnetic susceptibility is a function of external magnetic field, the dynamic exchange coupling length can vary with the sweeping of the external magnetic field. When these techniques are applied to a ferromagnet thicker than the dynamic exchange coupling length, special caution should be taken in the analysis of the spin-orbit torques. On the other hand, the short exchange coupling length may allow the second-harmonic method to probe spin-orbit torque at an individual surface, similarly to the anomalous spin-orbit torques observed via MOKE [23].

## Acknowledgement

The authors would like to express their great appreciation for Prof. Chia-ling Chien's inspirations on many scientists around the world. This work was supported by the National Science Foundation under Grant No. 2105218.

## Declaration of Competing Interest

The authors declare that they have no known competing financial interests.

## References

[1] S.O. Valenzuela, M. Tinkham, Direct electronic measurement of the spin Hall effect, Nature, 442 (2006) 176-179.
[2] E. Saitoh, M. Ueda, H. Miyajima, G. Tatara, Conversion of spin current into charge current at room temperature: Inverse spin-Hall effect, Applied physics letters, 88 (2006) 182509.

[3] D. MacNeill, G. Stiehl, M. Guimaraes, R. Buhrman, J. Park, D. Ralph, Control of spin–orbit torques through crystal symmetry in WTe2/ferromagnet bilayers, Nature Physics, 13 (2017) 300-305.

[4] A.M. Humphries, T. Wang, E.R. Edwards, S.R. Allen, J.M. Shaw, H.T. Nembach, J.Q. Xiao, T.J. Silva, X. Fan, Observation of spin-orbit effects with spin rotation symmetry, Nature communications, 8 (2017) 911

[5] M. Kimata, H. Chen, K. Kondou, S. Sugimoto, P.K. Muduli, M. Ikhlas, Y. Omori, T. Tomita, A. MacDonald, S. Nakatsuji, Magnetic and magnetic inverse spin Hall effects in a non-collinear antiferromagnet, Nature, 565 (2019) 627-630.

[6] I.M. Miron, K. Garello, G. Gaudin, P.-J. Zermatten, M.V. Costache, S. Auffret, S. Bandiera, B. Rodmacq, A. Schuhl, P. Gambardella, Perpendicular switching of a single ferromagnetic layer induced by in-plane current injection, Nature, 476 (2011) 189-193.

[7] L. Liu, C.-F. Pai, Y. Li, H. Tseng, D. Ralph, R. Buhrman, Spin-torque switching with the giant spin Hall effect of tantalum, Science, 336 (2012) 555-558.

[8] I.M. Miron, T. Moore, H. Szambolics, L.D. Buda-Prejbeanu, S. Auffret, B. Rodmacq, S. Pizzini, J. Vogel, M. Bonfim, A. Schuhl, Fast current-induced domain-wall motion controlled by the Rashba effect, Nature materials, 10 (2011) 419-423.

[9] S. Emori, U. Bauer, S.-M. Ahn, E. Martinez, G.S. Beach, Current-driven dynamics of chiral ferromagnetic domain walls, Nature materials, 12 (2013) 611-616.

[10] W. Jiang, P. Upadhyaya, W. Zhang, G. Yu, M.B. Jungfleisch, F.Y. Fradin, J.E. Pearson, Y. Tserkovnyak, K.L. Wang, O. Heinonen, Blowing magnetic skyrmion bubbles, Science, 349 (2015) 283-286.

[11] S.-W. Lee, K.-J. Lee, Emerging three-terminal magnetic memory devices, Proceedings of the IEEE, 104 (2016) 1831-1843.

[12] L. Liu, T. Moriyama, D. Ralph, R. Buhrman, Spin-torque ferromagnetic resonance induced by the spin Hall effect, Physical review letters, 106 (2011) 036601.

[13] S. Karimeddiny, J.A. Mittelstaedt, R.A. Buhrman, D.C. Ralph, Transverse and longitudinal spin-torque ferromagnetic resonance for improved measurement of spin-orbit torque, Physical Review Applied, 14 (2020) 024024.

[14] J. Kim, J. Sinha, M. Hayashi, M. Yamanouchi, S. Fukami, T. Suzuki, S. Mitani, H. Ohno, Layer thickness dependence of the current-induced effective field vector in Ta| CoFeB| MgO, Nature materials, 12 (2013) 240-245.

[15] X. Fan, J. Wu, Y. Chen, M.J. Jerry, H. Zhang, J.Q. Xiao, Observation of the nonlocal spin-orbital effective field, Nature communications, 4 (2013) 1799.

[16] K. Garello, I.M. Miron, C.O. Avci, F. Freimuth, Y. Mokrousov, S. Blügel, S. Auffret, O. Boulle, G. Gaudin, P. Gambardella, Symmetry and magnitude of spin–orbit torques in ferromagnetic heterostructures, Nature nanotechnology, 8 (2013) 587-593.

[17] X. Fan, H. Celik, J. Wu, C. Ni, K.-J. Lee, V.O. Lorenz, J.Q. Xiao, Quantifying interface and bulk contributions to spin–orbit torque in magnetic bilayers, Nature communications, 5 (2014) 3042.

[18] X. Fan, A.R. Mellnik, W. Wang, N. Reynolds, T. Wang, H. Celik, V.O. Lorenz, D.C. Ralph, J.Q. Xiao, All-optical vector measurement of spin-orbit-induced torques using both polar and quadratic magneto-optic Kerr effects, Applied Physics Letters, 109 (2016) 122406.

[19] M.-H. Nguyen, C.-F. Pai, Spin–orbit torque characterization in a nutshell, APL Materials, 9 (2021) 030902.

[20] D.C. Ralph, M.D. Stiles, Spin transfer torques, Journal of Magnetism and Magnetic Materials, 320 (2008) 1190-1216.

[21] A. Davidson, V.P. Amin, W.S. Aljuaid, P.M. Haney, X. Fan, Perspectives of electrically generated spin currents in ferromagnetic materials, Physics Letters A, 384 (2020) 126228.

[22] A. Ghosh, S. Auffret, U. Ebels, W. Bailey, Penetration depth of transverse spin current in ultrathin ferromagnets, Physical review letters, 109 (2012) 127202.

[23] W. Wang, T. Wang, V.P. Amin, Y. Wang, A. Radhakrishnan, A. Davidson, S.R. Allen, T.J. Silva, H. Ohldag, D. Balzar, Anomalous spin–orbit torques in magnetic single-layer films, Nature nanotechnology, 14 (2019) 819-824.
[24] G.S. Abo, Y.-K. Hong, J. Park, J. Lee, W. Lee, B.-C. Choi, Definition of magnetic exchange length, IEEE Transactions on Magnetics, 49 (2013) 4937-4939.
[25] Z. Qiu, S.D. Bader, Surface magneto-optic Kerr effect (SMOKE), Journal of magnetism and magnetic materials, 200 (1999) 664-678.
[26] C.O. Avci, K. Garello, M. Gabureac, A. Ghosh, A. Fuhrer, S.F. Alvarado, P. Gambardella, Interplay of spin-orbit torque and thermoelectric effects in ferromagnet/normal-metal bilayers, Physical Review B, 90 (2014) 224427.
[27] E.H. Sondheimer, The mean free path of electrons in metals, Advances in physics, 50 (2001) 499-537.
[28] S.-Y. Huang, X. Fan, D. Qu, Y. Chen, W. Wang, J. Wu, T. Chen, J. Xiao, C. Chien, Transport magnetic proximity effects in platinum, Physical review letters, 109 (2012) 107204.
[29] H. Nakayama, M. Althammer, Y.-T. Chen, K.-i. Uchida, Y. Kajiwara, D. Kikuchi, T. Ohtani, S. Geprägs, M. Opel, S. Takahashi, Spin Hall magnetoresistance induced by a nonequilibrium proximity effect, Physical review letters, 110 (2013) 206601.

## Appendices

A. Propagation matrix method to compute the MOKE response

The MOKE response through the heterostructure is derived using the propagation method. The linearly polarized light is decomposed into a superposition of a left-handed and a right-handed circularly polarized light with corresponding index of refractions in the magnetic materials as $n_\text{L} = n + Qm_\text{z}$, and $n_\text{R} = n - Qm_\text{z}$ respectively. There is no difference in index of refractions for left-handed and right-handed circularly polarized lights in other nonmagnetic layers. We'll calculate the left-handed circularly polarized light as an example.

For the *i*-th layer, we consider the light propagating in the +*z* direction having an electric field described as $E_\text{i}^+ e^{jn_\text{i}\omega z/c}$, where $n_\text{i}$ is the index of refraction of the *i*-th layer, $\omega$ is the angular frequency, and *c* is the speed of light. Similarly, the electric field propagating in the -*z* direction is described as $E_\text{i}^- e^{-jn_\text{i}\omega z/c}$. At an interface between the *i*-th and (*i*+1)-th layers, assuming *z* = 0, the Maxwell's boundary conditions dictate that

$$\begin{cases} E_\text{i}^+ + E_\text{i}^- = E_\text{i+1}^+ + E_\text{i+1}^- \\ \dfrac{E_\text{i}^+ - E_\text{i}^-}{n_\text{i}} = \dfrac{E_\text{i+1}^+ - E_\text{i+1}^-}{n_\text{i+1}} \end{cases}, \tag{A1}$$

which can be written as a matrix form, $\begin{pmatrix} E_\text{i+1}^+ \\ E_\text{i+1}^- \end{pmatrix} = \begin{bmatrix} \dfrac{n_\text{i}+n_\text{i+1}}{2n_\text{i}} & \dfrac{n_\text{i}-n_\text{i+1}}{2n_\text{i}} \\ \dfrac{n_\text{i}-n_\text{i+1}}{2n_\text{i}} & \dfrac{n_\text{i}+n_\text{i+1}}{2n_\text{i}} \end{bmatrix} \begin{pmatrix} E_\text{i}^+ \\ E_\text{i}^- \end{pmatrix} = \Gamma(\text{i}, \text{i}+1) \begin{pmatrix} E_\text{i}^+ \\ E_\text{i}^- \end{pmatrix}$. Here the matrix $\Gamma(\text{i}, \text{i}+1)$ is a propagation matrix describing how the electric fields of the light change at the interface. Similarly, a propagation matrix describing how the electric fields change in the bulk of one uniform layer can be written as $T(i) = \begin{bmatrix} e^{jn_i\omega d/c} & 0 \\ 0 & e^{-jn_i\omega d/c} \end{bmatrix}$. Therefore, the initial incident light and the final transmitted light after the *n*-th layer can be related by the multiplication of a series of these propagation matrices,

$$\begin{pmatrix} E_{n+1}^+ \\ 0 \end{pmatrix} = [\prod_{i=1}^{n} \Gamma(i, i+1) \, T(i)] \Gamma(0,1) \begin{pmatrix} E_0^+ \\ E_0^- \end{pmatrix} = P \begin{pmatrix} E_0^+ \\ E_0^- \end{pmatrix}, \quad (A2)$$

where we use $P = \begin{bmatrix} P_{11}(n_L) & P_{12}(n_L) \\ P_{21}(n_L) & P_{22}(n_L) \end{bmatrix}$ to denote the total propagation matrix. It should be pointed out that each matrix element here is a function of $n_L$. When right-handed circularly polarized light is used, the matrix element should be changed by replacing $n_L$ with $n_R$.

Since the MOKE measurement measures reflection, i.e. $\frac{E_0^-}{E_0^+}$, the overall Kerr signal from the heterostructure can be derived as $\theta_{\text{kerr}} + j\kappa_{\text{kerr}} = j \frac{\frac{P_{21}(n_L)}{P_{22}(n_L)} - \frac{P_{21}(n_R)}{P_{22}(n_R)}}{\frac{P_{21}(n_L)}{P_{22}(n_L)} + \frac{P_{21}(n_R)}{P_{22}(n_R)}}$, where the real part $\theta_{\text{kerr}}$ is the Kerr rotation, and the imaginary part $\kappa_{\text{kerr}}$ is the Kerr ellipticity.

In the simulation, we use 780 nm for laser wavelength and the rest of parameters used are listed in the table below.

|  | n | Thickness (nm) | Q |
| --- | --- | --- | --- |
| Air | 1 | ∞ | NA |
| Py | 2.38 + 4.36j | varies | 0.0036 - 0.011j |
| Pt | 2.76 + 4.84j | 3 | NA |
| SiO$_2$ | 1.45 | 1000 | NA |
| Si wafer | 3.71 + 0.01j | ∞ | NA |

Table A1 Parameters used in the MOKE simulation.